\documentclass[twocolumn,printnumbers,amsmath,amssymb,showpacs,prl]{revtex4}
\usepackage{graphicx}
\usepackage{color}

\begin{document}

\title{Signatures of shear thinning$-$thickening transition in dense athermal shear flows}

\date{\today}

\author{Wen Zheng$^{1,2}$}
\author{Yu Shi$^1$}
\author{Ning Xu$^{1,*}$}
\affiliation{$^1$CAS Key Laboratory of Soft Matter Chemistry, Hefei National Laboratory for Physical Sciences at the Microscale $\&$ Department of Physics, University of Science and Technology of China, Hefei 230026, P. R. China; \\$^2$Department of Modern Mechanics, University of Science and Technology of China, Hefei 230026, P. R. China}

\begin{abstract}

In non-equilibrium molecular dynamics simulations of dense athermal shear flows, we observe the transition from shear thinning to shear thickening at a crossover shear rate $\dot\gamma_c$.  Shear thickening occurs when $\frac{{\rm d (ln} T_g)}{{\rm d (ln}\dot\gamma)}>2$ with $T_g$ the granular temperature.  At the transition, the pair distribution function shows the strongest anisotropy.  Meanwhile, the dynamics undergo apparent changes, signified by distinct scaling behaviors of the mean squared displacement and relaxation time on both sides of $\dot\gamma_c$.  These features serve as robust signatures of the shear thinning$-$thickening transition.

\end{abstract}


\maketitle

Complex fluids such as colloids, gels, emulsions, foams, and granular materials exhibit intriguing and complicated rheological phenomena subject to shear.  Unlike newtonian fluids whose shear viscosity is independent of the shear rate, complex fluids can behave shear thinning or thickening under proper conditions, characterized by the decrease or increase of the viscosity with increasing the shear rate \cite{vermant,wagner,cheng1,brown1}.  Shear thinning and thickening are attractive research topics with important applications \cite{wagner,lee}.  In practice, we always wish paints to shear thin, while shear thickening is desired in the design and manufacture of smart materials such as soft body armors.

The underlying mechanisms of shear thinning and thickening have been debated over decades.  Early simulations have suggested that the layering of particles along the direction of the shear flow contributes to shear thinning \cite{erpenbeck,heyes}, which has been suspected as an artifact of profile biased thermostat \cite{evans,delhommelle} and been challenged by most recent studies \cite{cheng2,xu_xinliang}.  A recent measure of the microscopic single-particle dynamics has proposed that shear thinning results from the decrease of the entropic force contribution \cite{cheng1}.  Besides, shear thinning can be boosted by the presence of yield stress which on the contrary impedes the emergence of shear thickening \cite{brown2}.  Compared to shear thinning, our understanding of shear thickening is even poorer, because fewer systems exhibit shear thickening and shear thickening usually happens at high shear rates where probing is difficult.  Multiple mechanisms, {\it e.g.}~the order-disorder transition and formation of hydroclusters \cite{wagner,hoffman,heyes}, have been proposed to explain shear thickening.  Recent studies have also unveiled possible links between shear thickening and jamming of constituent particles due to dilation \cite{brown3,brown4,waitukaitis,fall}.  Shear thickening has been observed in both Brownian and Non-Brownian suspensions \cite{cheng1,brown2,fall} and in simulations with and without taking hydrodynamics into account \cite{wagner,heyes,delhommelle,foss}.  It then remains an open question whether shear thickening originates from the same mechanism for various systems.

In this letter, we study the rheology of planar shear flows of athermal granular systems {\it via} non-equilibrium molecular dynamics simulations.  By applying the simple model without the interference of hydrodynamics and thermostat, we observe the transition from shear thinning to shear thickening.  This transition is accompanied with some robust signatures which may not be limited to athermal systems and may thus provide us with a general picture of shear thickening.

Our systems are two-dimensional $L\times L$ squares consisting of disks with an identical mass $m$.  A half of the disks have a diameter of $\sigma$, while the other half have $\sigma_L=1.4\sigma$.  Lees-Edwards boundary conditions \cite{allen} are applied with the shear being imposed in the $x-$direction.  SLLOD equations of motion assuming a linear velocity profile \cite{evans_book} are employed:
\begin{eqnarray}
\frac{{\rm d}\vec{r}_i}{{\rm d}t} & = &\vec{v}_i+\dot{\gamma} y_i \hat{x},  \label{rt}\\
\frac{{\rm d} \vec{v}_i}{{\rm d}t} & = & \frac{1}{m}\sum_j \vec{F}_{ij} - \dot{\gamma}{v_{yi}}\hat{x}, \label{vt}
\end{eqnarray}
where $\vec{r}_i=(x_i,y_i)$ and $\vec{v}_i=(v_{xi}, v_{yi})$ are the location and random velocity of particle $i$, $\dot{\gamma}=\frac{{\rm d}\gamma}{{\rm d}t}$ is the shear rate with $\gamma$ the shear strain, $\vec{F}_{ij}$ is the force acting on particle $i$ by particle $j$, and the sum is over all particles $j$ interacting with particle $i$.  The force $\vec{F}_{ij}$ includes two parts, the elastic force $\vec{F}_{ij}^e=- \nabla V_{ij}$ and damping force $\vec{F}_{ij}^v= - \xi m\left(\vec{v}_{ij} +\dot\gamma y_{ij}\hat{x}\right)$ \cite{xu}, where $V_{ij}$, $\vec{v}_{ij}$, and $y_{ij}$ are the interaction potential, relative random velocity, and $y-$distance between particles $i$ and $j$, and $\xi$ is the damping coefficient.  Weeks-Chandler-Anderson (WCA) potential, {\it i.e.}~repulsive Lennard-Jones potential, is applied: $V_{ij}=\frac{\epsilon}{72} \left[ \left(\frac{r_{ij}}{\sigma_{ij}}\right)^{12}-2\left(\frac{r_{ij}}{\sigma_{ij}}\right)^6+1\right]$ when the separation between particles $i$ and $j$, $r_{ij}$, is smaller than the sum of their radii $\sigma_{ij}$, and zero otherwise.  We use $\sigma$, $m$, and $\epsilon$ as the units of length, mass, and energy.  The units of temperature and time are $\epsilon/k_B$ and $\sigma/\sqrt{\epsilon / m}$, respectively, where $k_B$ is the Boltzmann constant.

We apply Gear predictor-corrector algorithm to integrate Eqs.~(\ref{rt}) and (\ref{vt}) at a constant packing fraction $\phi=0.85$ above the $T=0$ jamming transition at $\phi_c\approx 0.84$ \cite{liu,ohern,zhao,wang}.  The systems at rest are jammed solids obtained from L-BFGS energy minimization \cite{lbfgs}.  Data are collected and averaged over time after the systems have been sheared over a time much longer than the characteristic time scale $1/\dot{\gamma}$ and steady shear flows without the memory of their initial states are achieved.  The shear stress $\Sigma_{xy}$ is calculated from
\begin{eqnarray}
\Sigma_{xy} &=& \Sigma_{xy}^{id} + \Sigma_{xy}^{ex} \nonumber \\
&=& -\frac{m}{L^2}\sum_{i=1}^N v_{xi}v_{yi} - \frac{1}{L^2} \sum_{i=1}^{N-1} \sum_{j=i+1}^N x_{ij}F_{yij},
\end{eqnarray}
where $\Sigma_{xy}^{id}$ and $\Sigma_{xy}^{ex}$ are the ideal gas stress and excess stress from particle interactions.  Correspondingly, the shear viscosity can be divided into two parts:
\begin{equation}
\eta=\eta^{id}+\eta^{ex}=\Sigma_{xy}^{id} / \dot{\gamma} + \Sigma_{xy}^{ex} / \dot{\gamma}. \label{eta1}
\end{equation}
We vary the number of particles $N$ from $256$ to $4096$ to verify that our results do not show significant system size dependence.  Here we only show results for $N=1024$.

In a steady shear flow, energy injected into the flow by the shear force is dissipated by the damping force, from which we obtain
\begin{equation}
\Sigma_{xy} \dot{\gamma} L^2 = \xi m\sum_{(i,j)} \vec{v}_{ij}^2, \label{stress2}
\end{equation}
where the sum is over all interacting particle pairs.  Simple calculations of Eq.~({\ref{stress2}}) lead to the eddy viscosity \cite{evans}
\begin{equation}
\eta_e = \eta_{_{-2}}^T + \eta_{_{-2}}^C + \eta_{_{-1}} + \eta_{_0}, \label{eta2}
\end{equation}
where
\begin{eqnarray}
\eta_{_{-2}}^T  & = & \frac{\xi m}{L^2} \sum_{(i,j)} (\vec{v}_{i}^2+\vec{v}_{j}^2)\dot\gamma^{-2} \approx \frac{2\xi N z_b}{L^2} k_BT_g \dot\gamma^{-2}, \label{eta_2T}\\
\eta_{_{-2}}^C & = &  - \frac{2\xi m}{L^2} \sum_{(i,j)}\left(\vec{v}_{i}\cdot\vec{v}_{j}\right) \dot\gamma^{-2}, \label{eta_2C}\\
\eta_{_{-1}} & = & \frac{2\xi m}{L^2} \sum_{(i,j)} (y_{ij}v_{xij})\dot\gamma^{-1}, \label{eta_1}\\
\eta_{_0} & = & \frac{\xi m}{L^2} \sum_{(i,j)}y_{ij}^2. \label{eta_0}
\end{eqnarray}
On the right hand side of Eq.~(\ref{eta_2T}), $z_b$ is the average coordination number, and $T_g=\frac{m}{2Nk_B}\sum_{i=1}^N \vec{v}_{i}^2$ denotes the granular temperature, {\it i.e.}~effective temperature at short time scales \cite{berthier,xu}.  When introducing $T_g$ in Eq.~(\ref{eta_2T}), we assume that the random part of the kinetic energy, $\sum_{i=1}^N\frac{1}{2}m\vec{v}_{i}^2$, is equally partitioned to all particles.

The decomposition of the viscosity by Eqs.~(\ref{eta1}) and (\ref{eta2}) enables us to find out the actual source of the shear stress in control of the rheology.  Equation~(\ref{eta1}) is the microscopic expression of the viscosity from virial theorem, while Eq.~(\ref{eta2}) is straightforwardly derived from energy balance.  In homogeneous steady flows we would expect that $\eta=\eta_e$.

\begin{figure}
\vspace{0 in}
\includegraphics[width=0.485\textwidth]{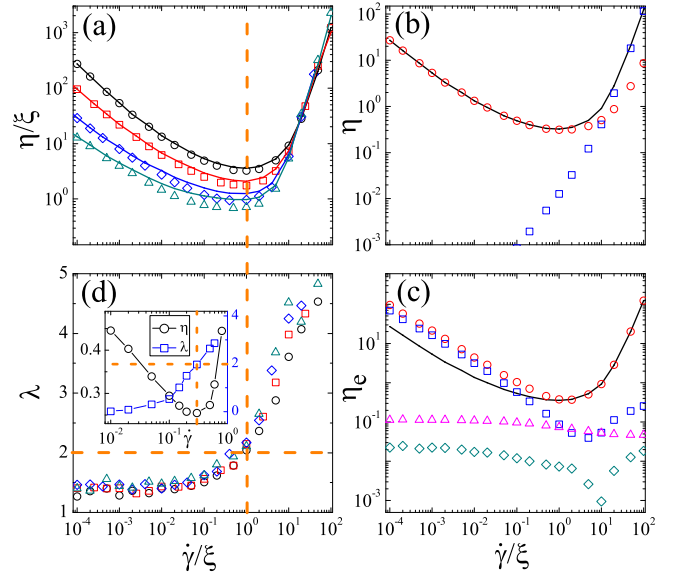}
\vspace{-0.2 in}
\caption{\label{fig:fig1} (color online) (a) Viscosity $\eta$ calculated from Eq.~(\ref{eta1}) against the shear rate $\dot\gamma$ at $\xi=0.1$ (black circles), $0.2$ (red squares), $0.5$ (blue diamonds), and $1.0$ (green triangles).  The solid curves are eddy viscosity $\eta_e$ calculated from Eq.~(\ref{eta2}).  (b) Shear rate dependence of the ideal gas viscosity $\eta^{id}$ (blue squares) and excess viscosity $\eta^{ex}$ (red circles) at $\xi=0.1$.  The black curve is the total viscosity.  (c) Shear rate dependence of the four components of the eddy viscosity, $\eta_{_{-2}}^T$ (red circles), $-\eta_{_{-2}}^C$ (blue squares), $|\eta_{_{-1}}|$ (green diamonds), and $\eta_{_0}$ (magenta triangles), at $\xi=0.1$.  The solid curve is the total eddy viscosity.  (d) Shear rate dependence of the rate of granular temperature increase $\lambda$ at the same damping coefficients as panel (a).  Inset: Comparison of $\eta$ and $\lambda$ for a thermostated system at $\frac{m}{Nk_B}\sum_iv_{yi}^2=0.1$.
}
\end{figure}

Figure~\ref{fig:fig1}(a) shows flow curves at four different damping coefficients ranging from $0.1$ to $1$.  There is a crossover shear rate, $\dot\gamma_c$, separating shear thinning at low shear rates from shear thickening at high shear rates.  Our systems at rest have a nonzero yield stress.  Shear thinning is consequently expected at low shear rates \cite{brown2}.  When scaling $\dot\gamma$ by $\xi$, we notice that $\dot\gamma_c\sim \xi$.  Scaling collapse at high shear rates in the shear thickening regime is also observed when $\eta/\xi$ is plotted against $\dot\gamma/\xi$, which approximately obeys a power law scaling $\eta\sim \dot\gamma^2$.  This scaling implies that the shear thickening flows here are not Bagnoldian ($\eta\sim \dot\gamma$) \cite{fall,bagnold}.  To check if our results are due to the Lees-Edwards boundary conditions, we have also studied systems confined between two walls at constant volume with the top wall moving in the $x-$direction at a constant speed and observed similar results.  As will be discussed in the following, our shear thickening flows are associated with high temperature and high pressure gas-like states induced by very high shear rates, which are in the same regime as previous simulations \cite{evans,delhommelle} but in different regime from recently reported Bagnoldian shear thickening flows arising from unjamming$-$jamming transition at low shear rates \cite{fall}.

As compared in Fig.~\ref{fig:fig1}(a), $\eta\approx \eta_e$ as expected, except in the vicinity of $\dot\gamma_c$ where $\Delta\eta=\eta_e-\eta>0$.  $\Delta\eta$ tends to increase with increasing $\xi$.  We have verified that $\Delta\eta>0$ is not a transient behavior by observing no time evolution.  In the shear rate regime where $\Delta\eta$ is apparently greater than $0$, the kinetics show visible anisotropy, {\it e.g.}~$\sum_i v_{xi}^2>\sum_i v_{yi}^2$.  As will be discussed, there exists structure anisotropy in the same regime as well.  The anisotropy should thus account for the nonzero $\Delta\eta$.  Although $\eta$ and $\eta_e$ are not completely equal, they exhibit the same $\dot\gamma_c$, so this inequality does not affect our discussions about the shear thinning$-$thickening transition using Eqs.~(\ref{eta1}) and (\ref{eta2}) respectively.

Let us first see what Eq.~(\ref{eta1}) tells us.  In Fig.~\ref{fig:fig1}(b), we compare $\eta^{id}$ with $\eta^{ex}$.  At low shear rates, the random motion of particles is so slow that $\eta^{ex}\gg\eta^{id}$.  With increasing the shear rate, $\eta^{id}$ grows up and eventually beats $\eta^{ex}$.  $\eta^{id}=\eta^{ex}$ at a crossover shear rate $\dot\gamma_g$.  When $\dot\gamma>\dot\gamma_g$, the system behaves effectively as a high temperature and high pressure gas with $\xi^{-1}$ and $\dot\gamma_c^{-1}$ the dominant time scales, which may be the cause of the high shear rate scaling collapse shown in Fig.~\ref{fig:fig1}(a).   Because $\dot\gamma_g>\dot\gamma_c$, Eq.~(\ref{eta1}) does not give us any clue about how the shear thinning$-$thickening transition happens.

In contrast, Eq.~(\ref{eta2}) is significantly useful to reveal the possible source of shear thickening.  In Fig.~\ref{fig:fig1}(c), we show all the four viscosity components in Eq.~(\ref{eta2}).  In the whole shear rate regime studied here, $\eta_{_0}$ and $\eta_{_{-1}}$ are small and negligible, $\eta_{_{-2}}^T>0$, and $\eta_{_{-2}}^C<0$.  At low shear rates, $|\eta_{_{-2}}^C|$ is smaller than but comparable to $\eta_{_{-2}}^T$, and shows similar shear rate dependence to $\eta_{_{-2}}^T$.  Near $\dot\gamma_c$, however, $\eta_{_{-2}}^T$ becomes significantly larger than $|\eta_{_{-2}}^C|$.  Shear thickening is thus mainly determined by $\eta_{_{-2}}^T$.  Assuming that the average coordination number $z_b$ does not vary largely with the shear rate, which is actually true for our systems, Eq.~(\ref{eta_2T}) implies that if the granular temperature $T_g$ varies faster than $\dot\gamma^2$ shear thickening would occur.

In athermal shear flows with a constant damping coefficient, $T_g$ increases with increasing the shear rate.  To estimate the rate of $T_g$ increase, we define a quantity $\lambda=\frac{{\rm d (ln} T_g)}{{\rm d (ln}\dot\gamma)}$ and plot it against the shear rate in Fig.~\ref{fig:fig1}(d).  At low shear rates, $\lambda\approx 1.5$, indicating that $T_g\sim \dot\gamma^{1.5}$.  This scaling law stops working near $\dot\gamma_c$.  When $\dot\gamma>\dot\gamma_c$, $\lambda>2$ and shear thickening happens, exactly as expected from our analysis of Eq.~(\ref{eta2}).

We have now established a direct link between the granular temperature and shear thickening for our model systems.  Is this link general or only specific to our systems?  Does it depend on particle interactions?  First of all, we have verified that systems with harmonic, Hertzian, and Lennard-Jones interactions come to the same conclusion. Next, let us see if $\lambda>2$ is also the condition of shear thickening in other typical model systems.

\begin{figure}
\vspace{0 in}
\includegraphics[width=0.485\textwidth]{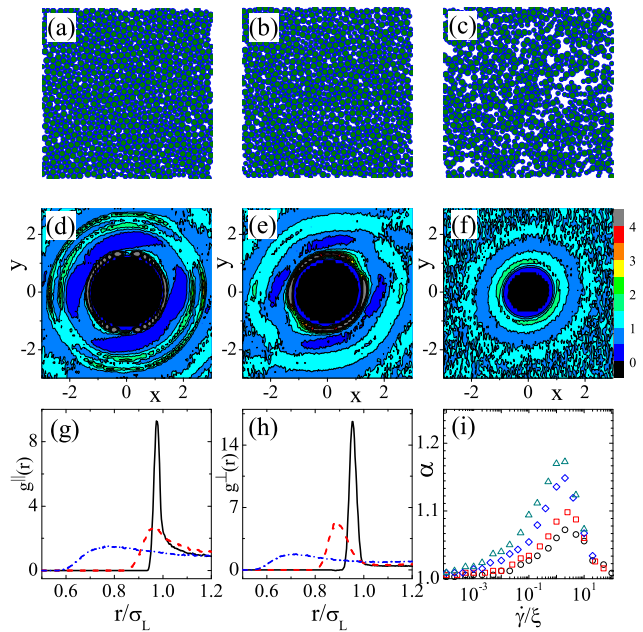}
\vspace{-0.2 in}
\caption{\label{fig:fig2} (color online) (a)$-$(c) Snapshots at $\xi=0.1$ and $\dot\gamma=0.01$, $0.1$, and $1$ with $\dot\gamma_c\approx 0.1$.  (d)$-$(f) Pair distribution functions $g(x,y)$ for systems shown in (a)$-$(c).  The value of $g(x,y)$ is quantified by the color. (g)$-$(h) Pair distribution functions parallel and perpendicular to $y=x$, $g^{\it \parallel}(r)$ and $g^{\perp}(r)$, at $\dot\gamma=0.01$ (black solid), $0.1$ (red dashed), and $1$ (blue dot-dashed).  (i) Anisotropy of $g(\vec{r})$ measured by $\alpha$ at $\xi=0.1$ (black circles), $0.2$ (red squares), $0.5$ (blue diamonds), and $1.0$ (green triangles).
}
\end{figure}

Besides our athermal model, typical models to study shear flows include the foam model \cite{durian,olsson}, Langevin dynamics \cite{ikeda,hansen}, and thermostated shear flows \cite{erpenbeck,evans,delhommelle,berthier}.  For the foam model described by the equation of motion, $\zeta \vec{v}_i=\sum_j \vec{F}_{ij}^e$, where $\zeta$ is equivalent to $\xi m$ in our model, simple calculations result in $\eta_e\sim \sum_i \vec{v}_i^2\dot\gamma^{-2}$.  We can find at once that shear thickening happens when $\frac{{\rm d} ({\rm ln}\sum_i \vec{v}_i^2)}{{\rm d}({\rm ln}\dot\gamma)}>2$, equivalent to $\lambda>2$ except that particle inertia is ignored.  When inertia is not neglected, $\lambda>2$ can be straightforwardly obtained as well.  For Langevin dynamics, a random force $\vec{R}_i(t)$ is added to the equation of motion for the foam model.  Because $\vec{v}_i$ and $\vec{R}_i$ are uncorrelated \cite{hansen}, we would expect the same result.  For thermostated shear flows, we have already known that when $T_g$ is fixed shear thickening cannot happen \cite{erpenbeck}, no matter if it is an artifact of thermostat \cite{allen}.  Even with more realistic thermostat, it has been reported that $T_g$ is much higher than the bath temperature \cite{allen}.  These previous observations may hint that shear thickening is related to the granular temperature in thermostated systems.  In the inset to Fig.~\ref{fig:fig1}(d), we show both $\eta$ and $\lambda$ against the shear rate for a thermostated shear flow governed by $\frac{{\rm d} \vec{v}_i}{{\rm d}t} =  \frac{1}{m}\sum_j \vec{F}_{ij}^e - \dot{\gamma}{v_{yi}}\hat{x} - \alpha v_{yi}\hat{y}$, where $\alpha$ maintains a constant kinetic energy in the $y-$direction \cite{berthier}.  For this model, energy balance results in $\eta_e\sim \alpha\sum_i v_{yi}^2 \dot\gamma^{-2}$, from which $\lambda>2$ cannot be recognized at all.  Interestingly, $\lambda>2$ still signifies shear thickening.

For foam model and Langevin dynamics, $\lambda>2$ is a straightforward consequence of energy balance for shear thickening to occur.  For our athermal model, this consequence is not so obvious because the precondition is that $\eta_{_{-2}}^T$ is significantly larger than the other three terms in Eq.~(\ref{eta2}), which turns out to be true.  For thermostated shear flows, $\lambda>2$ is totally unpredictable.  We thus propose that $\lambda>2$ is a generic signature of shear thickening at high shear rates.

The granular temperature is difficult to measure experimentally.  It is then interesting to know if there are any experimentally accessible signatures associated with the shear thinning$-$thickening transition.  As illustrated in panels (a)$-$(c) of Fig.~\ref{fig:fig2}, the system undergoes apparent changes from shear thinning to shear thickening.  In the shear thinning regime, the system is roughly uniform in space, while in the shear thickening regime particles tend to form instantaneous clusters with large overlaps and leave behind more voids.  We are thus inspired to search for possible structural or dynamical signatures of the shear thinning$-$thickening transition.

Recent studies have attempted to build the bridge between shear rheology and microstructure \cite{cheng2,nosenko,koumakis,dullens}. Here we measure the pair distribution function for large particles, $g(\vec{r})=g(x,y)=\frac{L^2}{(N / 2)^2}\left< \sum_i\sum_{j\ne i}\delta(\vec{r}(x,y)-\vec{r}_{ij})\right>$, with special attention to its anisotropy, where the sums are over all large particles and $\left< .\right>$ denotes the time average.  The contour plots of $g(\vec{r})$ shown in panels (d)$-$(f) of Fig.~\ref{fig:fig2} indicate that $g(\vec{r})$ is asymmetric in the vicinity of the shear thinning$-$thickening transition: the contours are elongated in the direction of $y=x$ and compressed in the direction perpendicular to $y=x$.

In order to quantify the asymmetry, we show in panels (g) and (h) of Fig.~\ref{fig:fig2} the pair distribution function parallel and perpendicular to $y=x$, denoted as $g^{\it \parallel}(r)$ and $g^{\perp}(r)$, respectively.  The first peak of $g^{\perp}(r)$ is higher than that of $g^{\it \parallel}(r)$.  Moreover, $r_s^{\perp}<r_s^{\parallel}$, where $r_s^{\perp}$ and $r_s^{\parallel}$ denotes the length at which $g^{\perp}(r)$ and $g^{\it \parallel}(r)$ start to be greater than zero.  These features imply that when the structure is asymmetric particles tend to have larger and more uniform overlaps in the direction perpendicular to $y=x$.  We define $\alpha=r_s^{\parallel}/r_s^{\perp}$ to characterize the asymmetry.  Figure~\ref{fig:fig2}(i) shows that with the increase of shear rate $\alpha$ first increases and then drops after a maximum.  The maximum emerges approximately at $\dot\gamma_c$.  Shear thinning$-$thickening transition is thus signified by the most pronounced structure anisotropy.

\begin{figure}
\vspace{0in}
\includegraphics[width=0.485\textwidth]{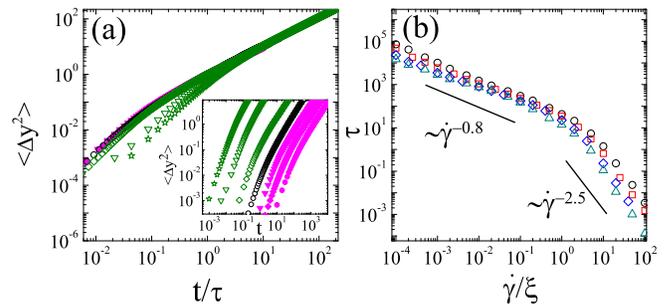}
\vspace{-0.2 in}
\caption{\label{fig:fig3} (color online) (a) Mean squared displacement $\left<\Delta y^2\right>$ in the shear thinning (magenta solid symbols) and thickening (green empty symbols) regimes at $\xi=0.1$. The black circles are at $\dot\gamma\approx\dot\gamma_c$.  The shear rate increases from the right to the left in the inset.  In the main panel, $t$ is scaled by the relaxation time $\tau$. (b) Shear rate dependence of the relaxation time $\tau$ at $\xi=0.1$ (black circles), $0.2$ (red squares), $0.5$ (blue diamonds), and $1.0$ (green triangles). The solid lines show power law behaviors in the shear thinning and thickening regimes.
}
\end{figure}

In the inset to Fig.~\ref{fig:fig3}(a), we show the $y-$component of the mean squared displacement (MSD) for large particles, $\left< \Delta y^2\right>$, where $\left< .\right>$ denotes the particle and ensemble average.  There is a short time ballistic motion ($\left< \Delta y^2\right>\sim t^2$) followed by diffusion ($\left< \Delta y^2\right>\sim t$) at long times.  Apparently, the relaxation time $\tau$ decreases with increasing the shear rate.  When we plot the MSD's against $t/\tau$, where $\tau$ is determined from $\left< \Delta y^2(\tau)\right>\approx 10\sigma^2$, as shown in Fig.~\ref{fig:fig3}(a), the MSD's in the shear thinning regime collapse nicely onto the same curve, while the ballistic parts of the shear thickening curves are still apart.  The scaling collapse of shear thinning curves implies that particles move ballistically to approximately the same distance.  In the shear thickening regime, however, particles move ballistically to a longer distance with increasing the shear rate.  As shown in Fig.~\ref{fig:fig2}(c), particles form instantaneous clusters in shear thickening flows and leave lots of voids.  If clustered particles move collectively at short times, the voids allow the particles to move ballistically to a longer distance, which may be the picture of the short time dynamics of shear thickening flows.

The distinct dynamics between shear thinning and thickening flows can also be identified from the shear rate evolution of the relaxation time.  As shown in Fig.~\ref{fig:fig3}(b), $\tau\sim \dot\gamma^{-0.8}$ in the shear thinning regime, in agreement with a recent experiment \cite{besseling}, while $\tau\sim \dot\gamma^{-2.5}$ in the shear thickening regime.  The distinct shear rate scaling of the relaxation time together with the breakdown of the scaling collapse of short time MSD's are thus dynamical signatures of the shear thinning$-$thickening transition.

In summary, in dense athermal shear flows, we observe robust signatures of the shear thinning$-$ thickening transition in the granular temperature, structure, and dynamics.  Preliminary results indicate that our findings may be general to various systems.  The quantities sorted out through this work would be useful and accessible tools in experiments and systematic simulations to understand shear thickening at high shear rates.

This work is supported by National Natural Science Foundation of China No.~11074228, National Basic Research Program of China (973 Program) No.~2012CB821500, CAS 100-Talent Program No.~2030020004, and Fundamental Research Funds for the Central Universities No.~2340000034.

\end{document}